\documentclass[a4paper,11pt]{article}
\usepackage{jheppub} 
\usepackage{subfigure}
\usepackage{color}
\usepackage{amssymb}
\usepackage{amsmath}
\usepackage{graphicx}
\usepackage{dcolumn}
\usepackage{bm}
\usepackage{bbm}
\usepackage{epsfig}
\usepackage{slashed}
\usepackage{amssymb}

\def\cot{\rm cot}

\def\ra{\rightarrow}

\def\bc{\begin{center}}
\def\ec{\end{center}}
\def\be{\begin{equation}}
\def\ee{\end{equation}}
\newcommand{\bear}{\begin{eqnarray}}
\newcommand{\eear}{\end{eqnarray}}
\newcommand{\ba}{\begin{array}}
\newcommand{\ea}{\end{array}}
\def\lsim{\mathrel{\rlap{\lower4pt\hbox{\hskip1pt$\sim$}}
    \raise1pt\hbox{$<$}}}         
\def\gsim{\mathrel{\rlap{\lower4pt\hbox{\hskip1pt$\sim$}}
    \raise1pt\hbox{$>$}}}

\def\bi{\begin{itemize}}
\def\ei{\end{itemize}}

\def\gtap{\raisebox{-.4ex}{\rlap{$\sim$}} \raisebox{.4ex}{$>$}}

\def\beq{\begin{displaymath}}
\def\eeq{\end{displaymath}}

\newcommand{\GeV}{~{\rm GeV}}
\newcommand{\TeV}{~{\rm TeV}}

\title{
Dilepton + jet signature of Split-UED at the LHC}
\author{Swarup Kumar Majee}
\author{and Seong Chan Park\footnote{s.park@skku.edu}}
\affiliation{Department of Physics and BK21 Physics Research Division, \\ Sungkyunkwan University, Suwon 440-746, Korea}

\abstract{We study the signature of dilepton and a hard jet ($\ell^+\ell^-j$) via heavy new gauge boson production in split universal extra dimension scenario where the Kaluza-Klein parity is conserved but the Kaluza-Klein number is not. A hard cut to the jet energy effectively removes virtually all possible backgrounds and provides a handle to search of new physics involving new neutral heavy states as the Kaluza-Klein $Z$ boson. The signature can be more generically used in search of other new states such as graviton and radion in warped extra dimension models. }

\begin{document} 
\maketitle
\flushbottom

\section{Introduction}
\label{sec:intro}

The robustness of the Standard Model (SM) of particle physics become
even more clearer as the CERN Large Hadron Collider (LHC) experiment continued
to have more data. The LHC should be already regarded as a successful experiment with its recent discovery of the Higgs boson. In its long time run, however, it aims besides Higgs boson search to explore new physics beyond the standard model. It is extremely important now to have some strategy to cover a wide range of new physics models but still have reasonably good handles to discriminate new physics signatures from the standard model backgrounds. 

We think that the signature with dilepton plus a hard jet
($\ell^+\ell^- j$) can provide a good chance to probe new physics
models with a new heavy (bosonic) state, $Z'$,  beyond the standard
$Z$-boson, for which putting a rather high cut in the jet energy can
efficiently remove the SM background mainly coming from the $Z$-boson
contributions. The heavy neutral state, $Z'$, can be found in many new
physics models such as models with extended gauge sectors (e.g.,
LR-model \cite{Pati:1974yy, Mohapatra:1974hk,
  Mohapatra:1974gc, Senjanovic:1975rk}, Little Higgs model
\cite{Giudice:2007fh}) and/or more spacetime dimensions and even with
heavy gravitational sectors (e.g., ADD  \cite{add1, anto, add2}, RS \cite{rs1,rs2}, UED \cite{acd} models). Having such a large variety, the expected collider phenomenology would also vary differently with the details of the setups but still some typical collider signatures can be studied provided that the mass of $Z'$ is in the reach of the LHC ($m_{Z'} \lsim 5$ TeV) and the coupling with the SM quarks and leptons is sizable. Hereafter, we would focus on $Z'$ in extra dimension models as an explicit example but the analysis can be applicable many different models too.

The extra dimensional models where the particles can access more extra space dimension than usual
$(3+1)$ space-time world, are quite promising as they can address several pressing problems in the SM.
The phenomenological introduction of the extra-dimensional model was
motivated to explain the well known hierarchy problem
\cite{hierarchy1,Gildener:1976ai,Susskind:1978ms,hierarchy2} allowing
gravity only to access some flat \cite{add1, anto, add2} or curved
\cite{rs1,rs2} space-time extra-dimensions. There are a few variations of
these models depending on whether gravity only, or gauge
bosons only or some other combinations of SM members are allowed to
access the extra dimensions. In this direction Universal Extra
Dimension (UED) is quite unique \cite{acd} due to the fact that $all$ the
SM-particles can access these new space dimensions.  

 In case of minimal UED (MUED) the extra space dimension is
to be compactified  
on an $S^1/\mathbb{Z}_2$ orbifold or equivalently on an interval, needed to have a chiral fermion at the
  zeroth level. Due to this orbifolding momentum along the fifth
  dimension, and hence the KK number, is no more
  conserved but a $\mathbb{Z}_2$-inversion symmetry about the middle point of the extra dimension, similar to R-symmetry in
  Minimal Supersymmetric Standard Model (MSSM), called {\em
    KK-parity} remains there. At the tree level, many production
  channels are restricted what allow to bring down the $R^{-1}$ upto a
  few hundred $\GeV$.  The model has a few
  good phenomenological implications, {\em like}, could be a natural
  source of dark matter particle \cite{Servant:2002aq,
    Matsumoto:2007dp, Park:2009cs, Chen:2009zs, Belanger:2010yx}, can
  give a TeV scale Grand Unified Theory \cite{Dienes:1998, Dienes:1998vg, Hossenfelder, Bhattacharyya:2006ym}, modify the supersymmetric Higgs boson mass \cite{
  Bhattacharyya:2007te}, and can be successfully explored at the CERN LHC
\cite{Macesanu:2002db, Macesanu:2002ew,Cheng:2002rn,
  Muck:2003kx,Rizzo:2001sd,Bhattacharyya:2009br,Kim:2011tq,
  Flacke:2012ke,Datta:2012tv, Datta:2012xy, Bandyopadhyay:2009gd, 
  Bhattacharyya:2005vm,Datta:2008,Bhattacherjee:2005,Bhattacherjee:2008,Ghosh:2012zc,Murayama:2011hj}. 

\begin{figure}[tbh]
  \centering
\includegraphics[width=0.5\textwidth
]{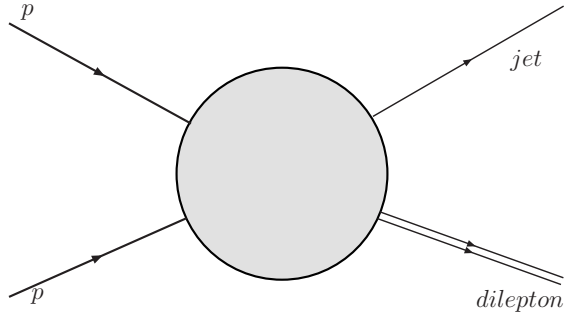}
\caption{\em Schematic diagram of jet $+$ dilepton production.}
  \label{fig:main}
\end{figure}
Because of its structural virtue all the SM fermions in this model can
have some vector like bulk mass terms, which is considered to be 
zero. However this bulk mass term is naturally included in split-UED
 scenario \cite{Park:2009cs,Chen:2009gz, Park:2010, Csaki:2010az, Huang:2012kz}. This mass term has a
great impact on the coupling and mass spectra, will be discussed in
the next section,  and the phonomenolgy is much different than the
MUED case. 
\begin{figure}[b]
{\hspace*{3mm}}  
\includegraphics[width=0.44\textwidth]{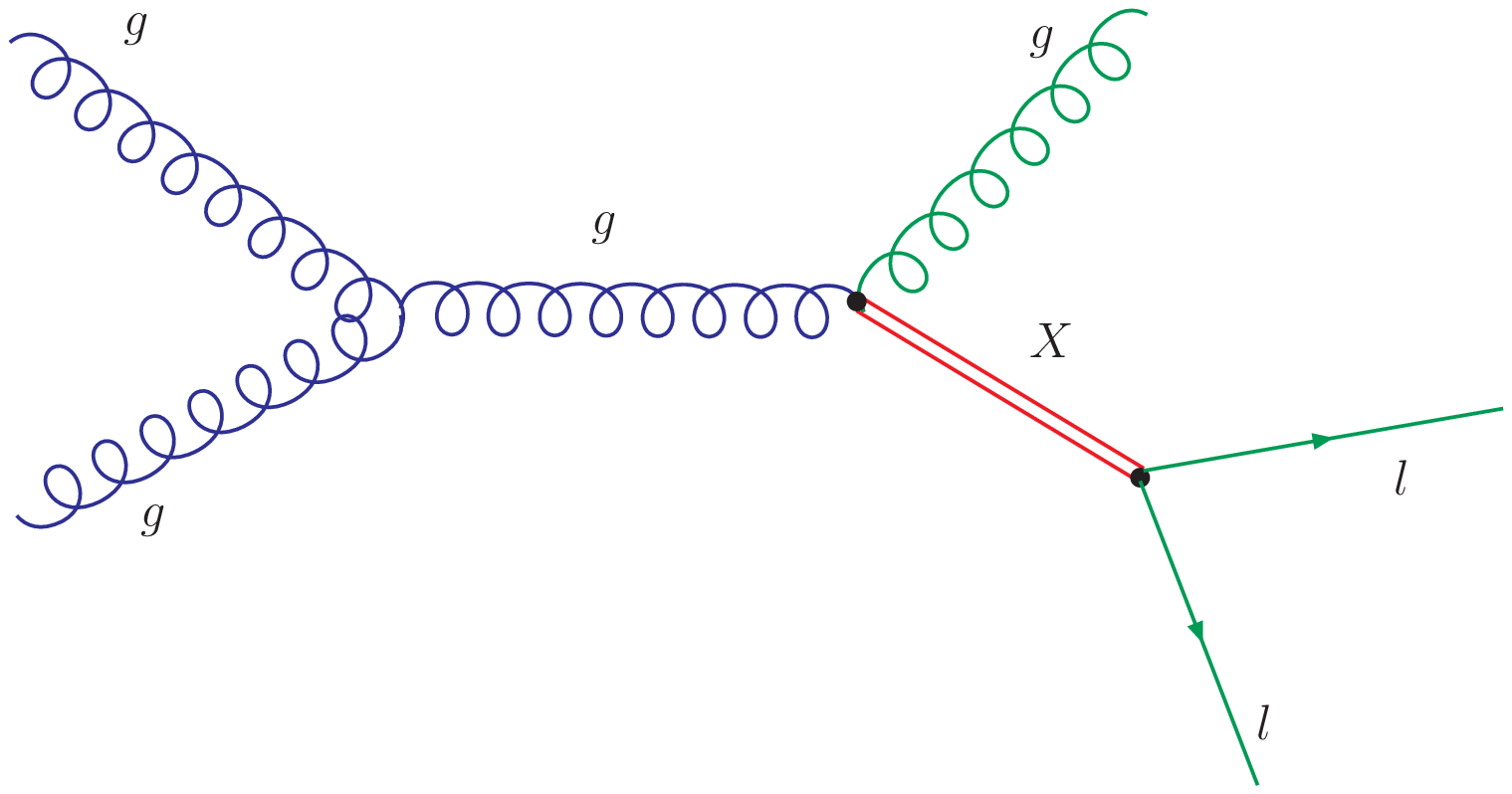}
{\hspace*{.4cm}} \includegraphics[width=0.44\textwidth]{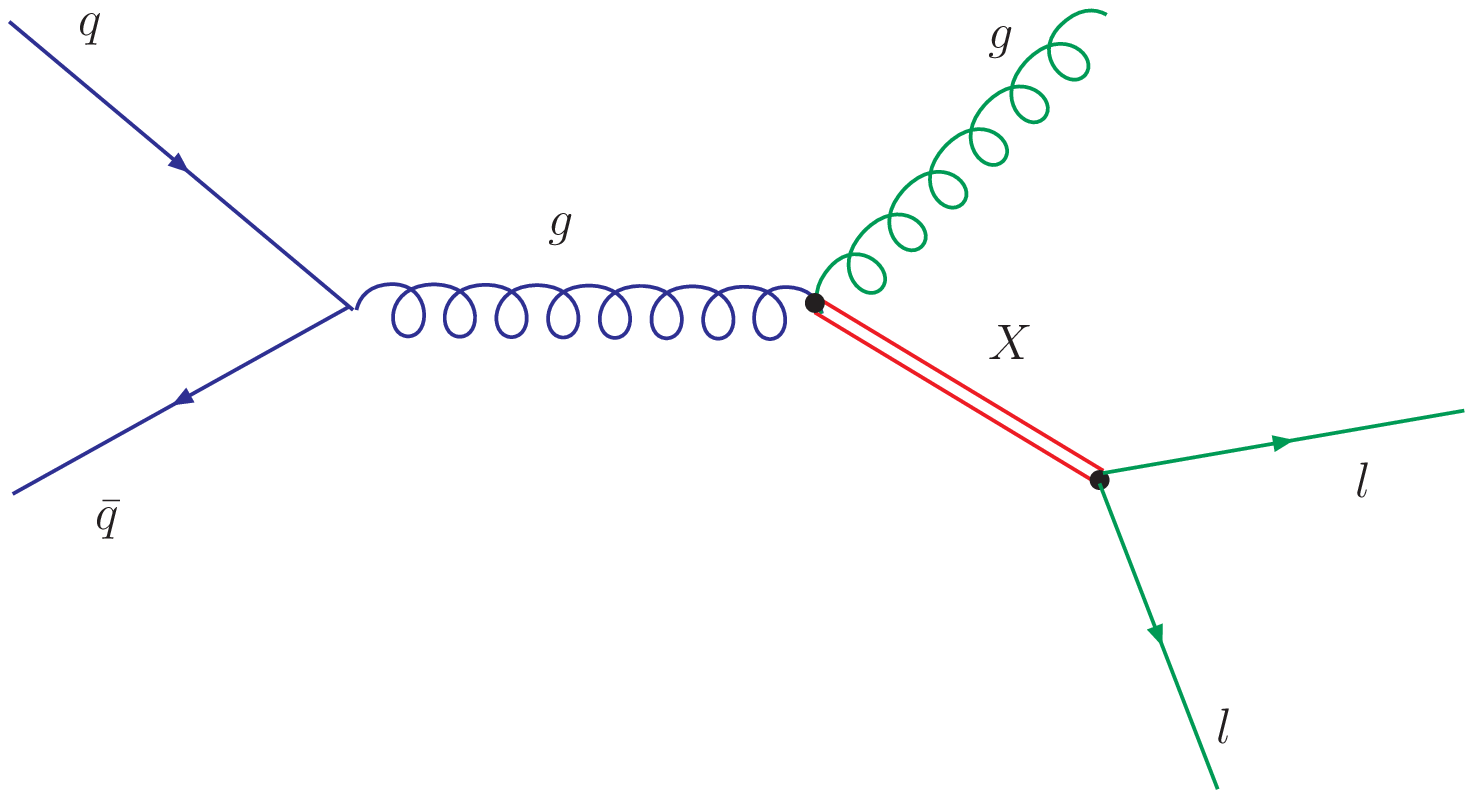}
\caption{\em RS model can give a similar final state, where Graviton
  will decay into two leptons.}
  \label{fig:rs}
\end{figure}
One interesting feature is that now we could have
a KK-number violating coupling at the tree level. As a matter of fact,
second KK-level weak gauge bosons can directly decay into a pair of
light fermions without any missing energy -- the scenario is absent in MUED, R-parity conserving MSSM or any other new physics
containing a $Z_2$-symmetry. We thus consider here this distinct 
channel $pp\to j\ell^+\ell^-$, as shown in Fig.-\ref{fig:main} to
probe split-UED at LHC. However, a single jet associated with dilepton
final state may also appear in some other new physics areas like
RS-model etc where, as shown in Fig.-\ref{fig:rs}, the dileptons are
produced from the decay mode of the Graviton, radion or some the other
particles. LHC has now reached to an energy scale either to discover or
turned down many new physics scenario. At this energy scale
it produce quite high energetic multiple number jets with hard
leptons. ATLAS \cite{Meade:2011zh} and CMS \cite{Nespolo:2011zz}
discuss the channel of $Z$ plus jets as a final state to test
perturbative QCD. It is thus timely to have a look at some new
physics in this directions. Here, we consider the split-UED scenario. 
We admit that the analysis we put here is at the parton level and so a
further detail analysis should be performed.

Collider signature of split-UED for different channel, where 
dilpetons are produced in $s-channel$ mediated by the $Z_2$ or
$\gamma_2$ has already been studied in
ref. \cite{Park:2009cs,Chen:2009gz, Kong:2010xk}. However, in this study we have a real production of the
second KK-excited mode of the weak gauge bosons associated with a
jet. The final state is different than the previous 
analysis. In this case, the jet will be produced {\em back-to-back} with
the heavy excited gauge boson  which will decay into
two leptons. Thus, all the final state particles,
the jet and two leptons, have large momentum. Due to this feature
of a single hard jet and two hard leptons we can easily use a strong
cut to make our final states free from any standard model
background. The analysis can probe upto a few TeV of $Z_2$ mass of the
model with $100~ {\rm fb}^{-1}$ luminosity at the LHC.

This paper is organised as follows. Briefly introducing the physics fo
split-Universal Extra Dimension in the next section we will write down
the possible wave function expansion of various gauge and fermion
fields in this model. In the next section, we then estimate the
possible mass spectra and their various branching ratios for the
related particles in our model.  We then discuss the production of the
second KK-mode of weak gauge bosons, which would decay either to zero
mode quarks or leptons, associated with a zero mode quark. With the
proper cut used here in this analysis, our jet $+$ dilepton final
states will be free from any sizable standard model background. We make our
conclusion in the last section.

\section{split-UED review}\label{review} 

In split-UED the full spacetime, $x^M = (x^0, x^1,x^2,x^3, y)$, include an extra dimension compactified on an $S^1/\mathbb{Z}_2$ orbifold, with a radius $R$. The orbifold can be regarded as an flat interval $[-L, L] = [-\pi R/2, \pi R/2]$ where the end points $y=\pm L$ correspond to the fixed points of the orbifold. All the SM fields, gauge bosons, Higgs and fermions, are in the full spacetime so that each of them can be decomposed into KK-states with a finite mass gap, $\Delta m \sim 1/R$.  The zero modes are identified with the SM field in the end.

One of the most interesting features of split-UED is the bulk mass terms for fermions. Indeed, the spinor in five dimensions includes both chiralities in four dimensional sense, so that it can have Dirac mass term, which is consistent with five dimensional Lorentz symmetry as well as gauge symmetries:
\begin{eqnarray}
{\cal L}_{\rm split-UED} = {\cal L}_{\rm MUED}+ \int_{-L}^L dy \, \sum_\Psi \left[M_\Psi(y) \overline{\Psi}(x^\mu,y) \Psi(x^\mu,y) + {\rm h.c.}\right],
\end{eqnarray}
where $\Psi(x^\mu, y) = \{Q, U^c, D^c, L, E^c\}$ denote the bulk fermion fields for quarks and leptons and $M_\Psi(y)=-M_\Psi(-y)$ is introduced to keep the KK-parity under which the spinor field transforms as $\overline{\Psi}\Psi(y) \to -\overline{\Psi}\Psi(-y)$. The bulk mass terms do not contribute to the masses for zero mode fermions, i.e., the SM fermions, but control the wave function profiles of each fermion along the extra dimension so that its interaction patterns with the KK gauge bosons could be modified.  

Here we simply collect some informations on KK masses and couplings in split-UED for our phenomenological study (the details of split-UED could be found in original Ref. e.g.,  \cite{Kong:2010xk}.):
\begin{itemize}
\item We are mostly interested in the KK-gauge boson couplings with the zero mode leptons and quarks, which can be obtained by 
\begin{eqnarray}
g_{Z_n \ell_0 \bar{\ell}_0}= g_{5D} \int_{-L}^L  \,dy f_{W_n} f_{\ell_0} f_{\ell_0} \equiv g_{\rm SM}{\cal F}_{00}^n(\mu_\ell L), \nonumber \\
g_{Z_n q_0 \bar{q}_0}= g_{5D} \int_{-L}^L  \,dy f_{W_n} f_{q_0} f_{q_0} \equiv g_{\rm SM}{\cal F}_{00}^n(\mu_q L), 
\end{eqnarray}
where $f_{q_0}$ and $f_{\ell_0}$ are the zero mode wave functions for the quark and lepton.The Standard Model couplings could be obtained by $g_{\rm SM} = g_5/\sqrt{2L}$. The 5D bulk mass parameters $\mu_q$ and $\mu_\ell$ are for quark and lepton, respectively.

\item Thanks to the KK-parity conservation, only KK-even modes have non-zero couplings.

\begin{eqnarray}
{\cal F}_{00}^{2n}(x)=\frac{x^2 \left[1-(-1)^n e^{2x}\right]\left[1-\coth\left(x\right)\right]}{\sqrt{2(1+\delta_{0n})}\left[x^2+n^2 \pi^2/4 \right]} \, . \label{eq:g002n2}
\label{eq:coupl}
\end{eqnarray}

\item The mass for the KK-$Z$-boson in split-UED is the same as in minimal-UED as there is no mass term allowed due to gauge symmetry:
\begin{eqnarray}
m_{Z^{(n)}} = \sqrt{m_Z^2 + \frac{\pi^2 n^2}{4L^2}}, \,\text{(split-UED)}
\end{eqnarray}
where $m_Z=91.1876\pm 0.0021 \GeV$ \cite{PDG} and the KK-number is positive integer, $n \in \mathbb{Z}^+$.

\item The KK mass for fermions in split-UED is  %
 \begin{eqnarray}
 m_{f^{(n)}}=  \sqrt{m_f^2 + \mu^2 + k^2_n},\,\text{(split-UED)}
  \end{eqnarray}
where $m_f = \lambda_f v$ is the mass obtained through Yukawa interaction and the Higgs vev.  The `fifth-momentum' $k_n$ is given as
 
\begin{eqnarray}
 \label{eq:kn}
k_n &&=\begin{cases}
\begin{cases}
i\kappa_1 & :\kappa_1=\kappa \in\left\{0<\kappa  ~ | ~\mu=-\kappa \coth \kappa L, \,\mu L<-1\right\} \\ 
k_1 & :k_1=k \in \left\{0 \leq k\leq \frac{\pi}{L} ~| ~\mu=-k \cot k L,\,\mu L\geq -1\right\} 
\end{cases} & \text{ for } n=1 \\ 
 \frac{n}{R}= \frac{n\pi}{2 L}& \text{ for } n=2,4,6,\cdots \\ 
k_n =k \in \left\{\frac{(n-2)\pi}{L}<k<\frac{(n-1) \pi}{L}  ~ | ~  \mu=-k \cot k L \right\} &\text{ for } n=3,5,7,\cdots \, .
\end{cases}
\end{eqnarray}

\end{itemize}

\section{Collider physics}
\subsection{Branching ratios}
In this paper we are interested to look into the production and decay
channel of the second KK-excited weak neutral gauge bosons $Z_2$ and
$\gamma_2$ associated with high energetic jet. Mixing of this heavy
gauge bosons with the corresponding zero modes are almost zero,
so the $\gamma_2$ is just the second KK excited mode of the $B$-boson
while $Z_2$ the corresponding excited mode of the third component of the
$W$-boson. The branching fractions certainly depend on the bulk mass
parameter $\mu$. A very small bulk mass term, $|\mu| << 1$, the
phenomenology would be similar as in UED, so we briefly mention here 
the possible branching ratios for the two heavy weak gauge bosons,
$\gamma_2$ and $Z_2$ for the region where $|\mu| ~\gtap ~ 0.5$.  To
discuss this, we consider three different cases -- where the bulk mass
parameter is universal to both quarks and leptons and two other
situations in which either quarks or leptons have zero bulk mass. In
this case, we do have tree level coupling, stronger than the SM
coupling constant, of $\gamma_2$ and $Z_2$ decaying into fermion pair,
which is suppressed by the loop level in MUED case. The decay widths
for the $\gamma_2$ and $Z_2$ decaying into a pair of the SM fermion
final states are roughly given by $\Gamma \sim \frac{M_{\gamma_2/Z_2
  }}{8\pi} (g_L^2 + g_R^2)$, where couplings $g_L$ and $g_R$ are
discussed below.

As the fermions get more radiative correction than $\gamma_n$, so
for $n=2$ its decay into either two first KK-mode fermion pair or a combination of a
zero mode fermion and second KK-mode of fermion is kinematically not
allowed. Thus this $\gamma_2$ state would completely decay into a pair
of standard fermion-antifermion. In the case of the universal bulk
mass term for all the quarks and leptons, the $\gamma_2-f_0-\bar{f_0}$
coupling $g_{L/R}$ is completely governed by the corresponding
hypercharge quantum number of the associated fermion and proportional
to the $U(1)$ coupling constant.  The branching ratio thus one can
easily read, as given in Table-\ref{table:BR} $36.7\%$ for the dijet,
$4.15\%$ for the $b\bar{b}$, $36.7\%$ for the $t\bar{t}$ and $25\%$
for the dilepton,  $12.5\%$ for $\tau \bar{\tau}$ and the rest $7.5\%$
into neutrino as missing energy.  

In case of $Z_2$, the gauge boson get some radiative corrections and
to give the branching ratios we will reiterate what is discussed
before in ref \cite{Kong:2010xk}. This radiative correction and the
zero mode electroweak mass makes $Z_2$ as heavy as roughly around 1.07
times that of the $\gamma_2$ mass \cite{radiative1, radiative2} . 
However, although the second
excited state of different fermions do not receive much corrections from the
bulk mass but the corresponding coupling constants $Z_2-f_0-f_2$ are
highly suppressed for a large value of $\mu$, and the same is also
true for the $Z_2-f_1-f_1$ coupling constant. Thus, the only dominant
decay mode left is decaying into a pair of SM fermions. As $Z_2$
is basically $W^3_2$, the corresponding decay mode is simply the weak coupling constants and hence $Z_2$ would decay into
left-handed particles only. For a universal bulk mass parameter, $Z_2$
dominantly decay into dijet $50\%$ and $12.5\%$ into both $b\bar{b}$
and $t\bar{t}$ while $8.3\%$ into dilepton, $4.2\%$ into
$\tau\bar{\tau}$ and $12.5\%$ into missing energies. 

Since the $Z_2$ and $\gamma_2$ couplings with the fermion pair solely
depend on the $\mu$-parameters, the corresponding leptons and quarks
channels are off once we consider the vanishing $\mu_q$ and $\mu_l$
respectively.

In the limit $\mu_q=0$ where vector bosons now only decay into the leptonic
final states the branching ratio for the $\gamma_2$ decay
mode is thus given as $55.6\%$ into dilepton,  $27.7\%$ into
$\tau\bar{\tau}$ pair and the rest $16.7\%$ into missing energies
while the same for the $Z_2$ are respectively given $33.3\%$, $16.7\%$
and $50\%$. On the other hand, In the limit $\mu_q=0$, both $Z_2$ and
$\gamma_2$ decay into dijet with $66.7\%$ branching ratio while the
same for the decay mode into $b\bar{b}$ and $t\bar{t}$
pair is $16.65\%$  each.  The branching ratios for the $\gamma_2$ decay
into $b\bar{b}$ and $t\bar{t}$ pair are respectively given by
$7.5\%$ and $25.8\%$, as shown in Table-\ref{table:BR}. As we can see
the branching ratios are varying in these two different scenarios, so
the corresponding collider phenomenology will be highly effected. This
will be discussed elsewhere in future analysis.

\begin{table}[tbh]
\begin{center}
\begin{tabular}{|c|c|c|c|c|c|c|} \hline

&\multicolumn{6}{|c|}{Branching Ratios (in $\%$ )} \\ \cline{2-7} 
& \multicolumn{2}{|c|}{$\mu_q = \mu_l$} & \multicolumn{2}{|c|}{$\mu_q = 0$} & \multicolumn{2}{|c|} {$\mu_l=0$} \\ \hline
Decay  &&&&&& \\
Channel & $Z_2 \ra f\bar{f}$ & $\gamma_2 \ra f\bar{f}$ & $Z_2 \ra f\bar{f}$ & $\gamma_2 \ra f\bar{f}$ & $Z_2 \ra f\bar{f}$ & $\gamma_2 \ra f\bar{f}$  \\ \hline 
&&&&&& \\ 
dijet &50&36.7& 0&0&66.7&66.7 \\ \hline
&&&&&& \\ 
$b\bar{b}$ &12.5&4.15&0&0&16.65&7.5 \\ \hline
&&&&&& \\ 
$t\bar{t}$ &12.5&14.15&0&0&16.65&25.8 \\ \hline
&&&&&& \\ 
dilepton &8.3&25&33.3&55.6&0&0 \\ \hline
&&&&&& \\ 
$\tau \bar{\tau}$  &4.2&12.5&16.7&27.7&0&0 \\ \hline
&&&&&& \\ 
neutrinos &12.5&7.5&50&16.7&0&0 \\ \hline
\end{tabular}
\caption{\sf \small Branching ratios for the $Z_2$ and
  $\gamma_2$ decay modes into pair of light fermions.}
\label{table:BR}
\end{center}
\end{table}

\subsection{Production and decay of heavy weak neutral gauge bosons}
\begin{figure}[b]
{\hspace*{3mm}}  
\includegraphics[width=0.41\textwidth]{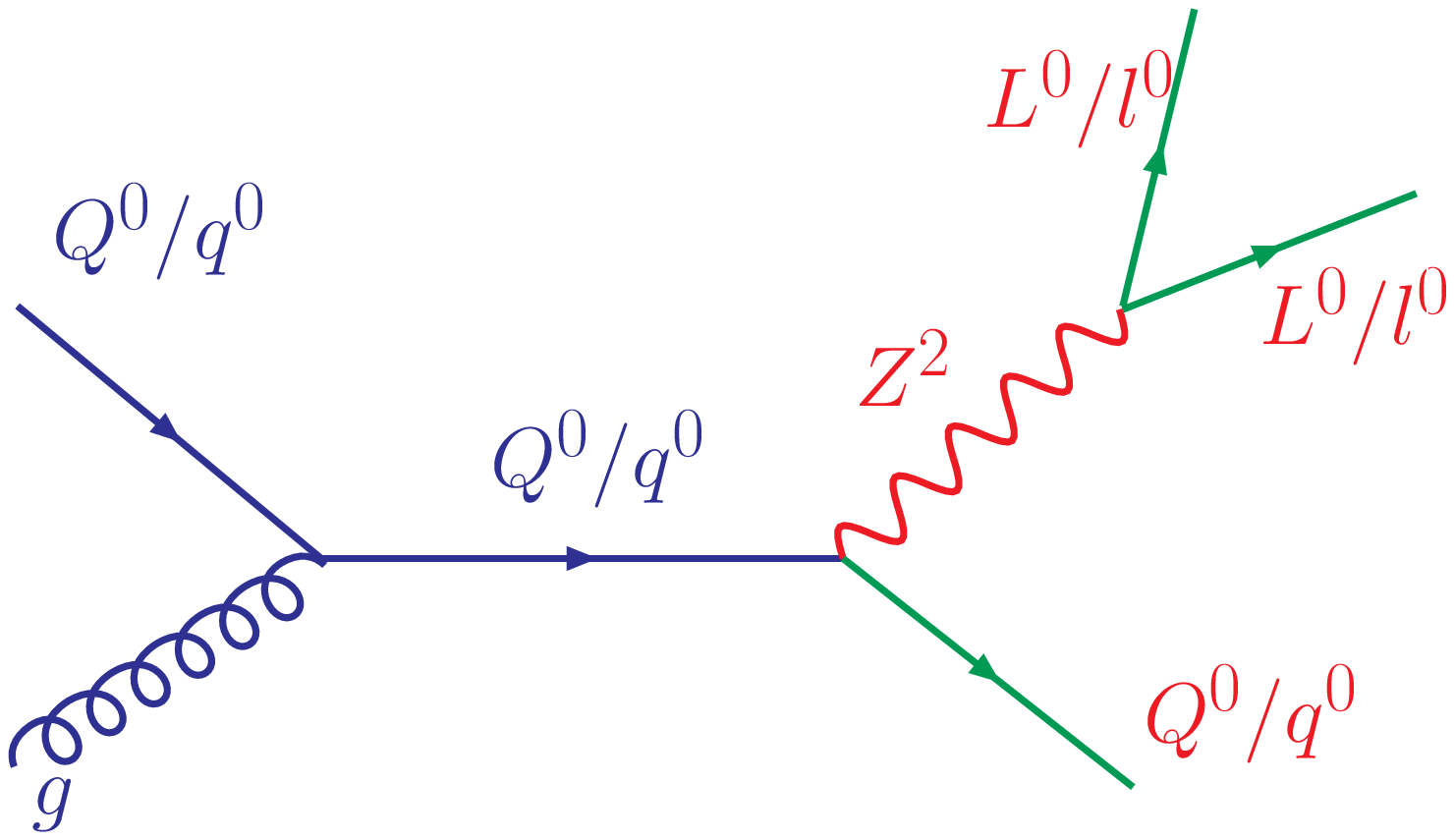}
{\hspace*{1.1cm}} \includegraphics[width=0.41\textwidth]{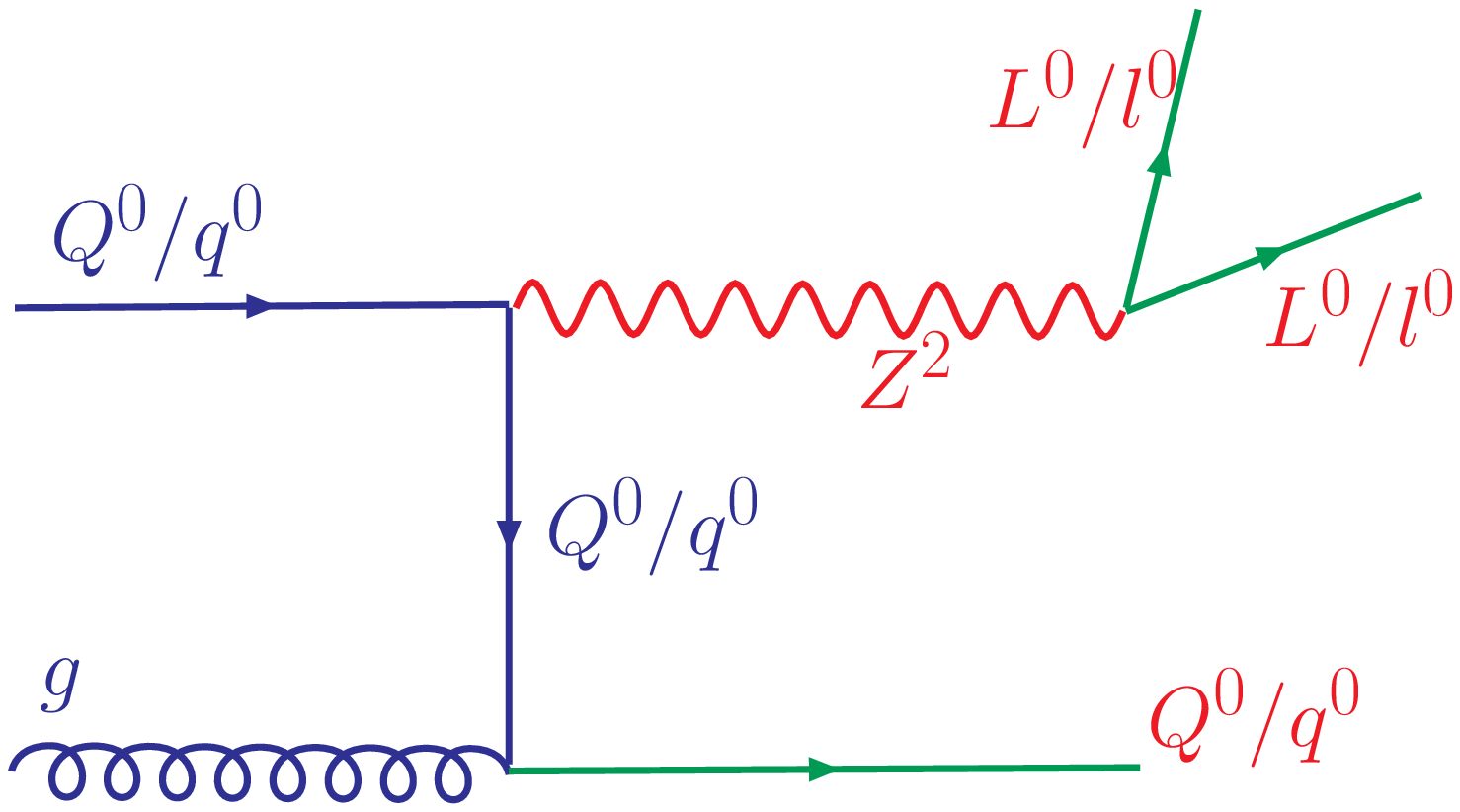}
\caption{\em Diagrams showing the production of a single jet $+$
  dilepton final state in split-UED model in s-channel (left) and in
  t-channel (right).}
  \label{f:split-prod}
\end{figure}
We will now consider the specific channel of one jet plus two leptons
final state for the split-UED model. The main production channel is $qg \ra
q^0Z^2({\rm or,} q^0\gamma^2)$, where $q^0$ could be both doublet as
well as singlet quark, as shown in Fig.-\ref{f:split-prod}. The
corresponding {\em s-} and {\em t-} channel parton level matrix
element square one can write as 

\begin{eqnarray}
|{\cal M}\left( qg \ra q^0Z^2/\gamma^2 \right) |^2 &=& 
 \frac{\pi  \alpha_s(\hat{s})}{3}
(g_L^2 + g_R^2) 
\left[- \frac{\hat s\hat t}{{\hat s}^2} - \frac{\hat s
      \hat t}{\hat t^2} + \frac{2\{(\hat s + \hat t)m_{Z^2/\gamma^2}^2 -
      m_{Z^2/\gamma^2}^4\}}{\hat s \hat t } \right]. \nonumber\\
\label{mainE}
\end{eqnarray}

In the above equation $\hat{s}$ and $\hat{t}$ are the parton level
Mandelstam variables for the above processes while  $m_{Z^2/\gamma^2}$
representing the mass of the final state weak vector boson
$Z^2/\gamma^2$. Since the mixing of the second KK-mode weak neutral
gauge bosons $Z^2/\gamma^2$ with the corresponding zero-th mode states
is almost negligible, as mentioned before $Z^2$ is basically the third
component of the second KK-mode of the $W$ boson while $\gamma^2$ is
just the same of the $B$-boson. So, the coupling $g_L$ is the only
nonzero coupling for the left-handed fermions in case of production
associated with the $Z_2$ boson. This is just the modified weak
$SU(2)$-weak gauge coupling, as discussed in section-\ref{review}. On
the other hand, in case of $\gamma^2$ the couplings $g_{L/R}$ will be
proportional to the corresponding left/right hypercharge quantum
number of the associated fermions times the modified $g_Y$ coupling constant. 
\begin{figure}[t]
  \centering
\includegraphics[width=0.6\textwidth
,height=0.6\textheight
,angle=-90
]{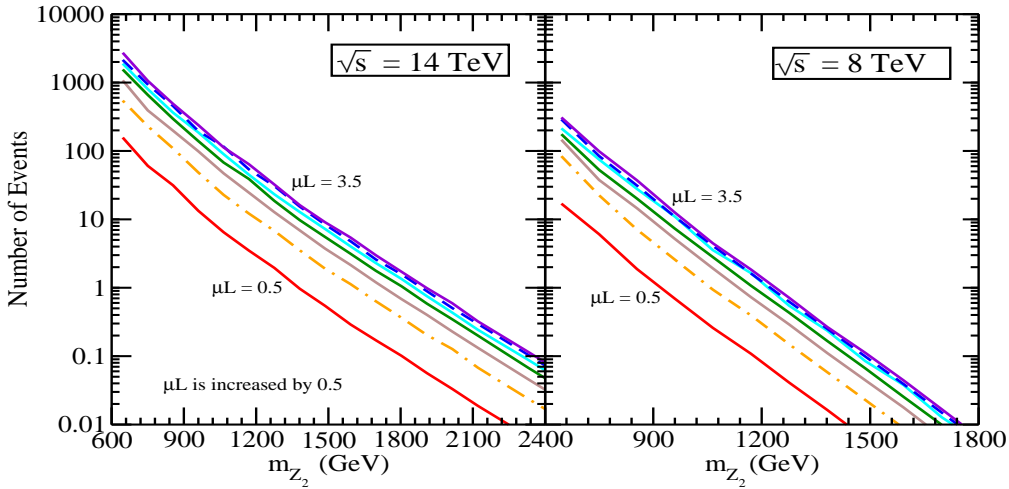}
\caption{\em  No. of signal events of  jet + dilepton channel.}
  \label{f:production}
\end{figure}
In our analysis, we have used CTEQ6 parton distribution \cite{Pumplin:2002vw}
with the renormalisation scale $Q^2 = \hat{s}$ to calculate the parton
level cross section at the LHC with center of mass $\sqrt{s}$ equals
to $14\TeV$ and $8\TeV$. We have shown the variation of the total
number of signal events for an integrated luminosity of
$100~\rm{fb}^{-1}$ with the inverse of compactification radius in
Fig.-\ref{f:production}. In this figure we have varied the bulk mass
parameter $\mu$ from 0.5 to 3.5, and we see the number of signal events
increases with the increase of the bulk mass parameter, as the
coupling constant increases while it decreases with the increase of
the $R^{-1}$ as the excited weak gauge bosons becoming heavier. 

Since the zeroth mode quark will be produced along with a heavy gauge
boson the jet will have a large momentum. In addition, these weak gauge
bosons will directly decay into a pair of charged leptons so they
would have large momentum as well. In this scenario we thus have a
hard jet and two hard leptons and {\em no} missing energy. This is
where the model differs from other new physics scenario, with an extra
$Z_2$-symmetry, where we can not have any tree level processes without
contributing to missing energy.  This high energetic jet and leptons
allow us to use a strong cut on the $p_T^{\rm jet}$ and the dilepton
invariant mass greater than $0.9$ times the $\gamma_2$ mass. In
addition, we have also used a rapidity cut $|\eta| < 2.5$ for leptons
and consider the leptons which are isolated from jet by $\Delta R >
0.7$ only. We have used Madgraph \cite{Maltoni:2002qb} to estimate the SM
background. The main channel which can contribute to a jet+ dilepton
backgrounds are $t\bar{t}$, $b\bar{b}$, $t\bar{b}$ and $b\bar{t}$
processes. We can also have some background from the Z $\gamma^*$
processes  or from a pair of Z-boson process. In either case the
dilepton may arise from the decay of one Z-boson or $\gamma$. On the
other hand another Z-boson will decay into a pair of jets. In case,
they are colse enough to fake as a single jet or only of the jet can
have large momentum. However, the above strong cuts are good enough to wash
out any of the mentioned background. 

\begin{figure}[t]
  \centering
\includegraphics[width=0.6\textwidth
,height=0.6\textheight
,angle=-90]{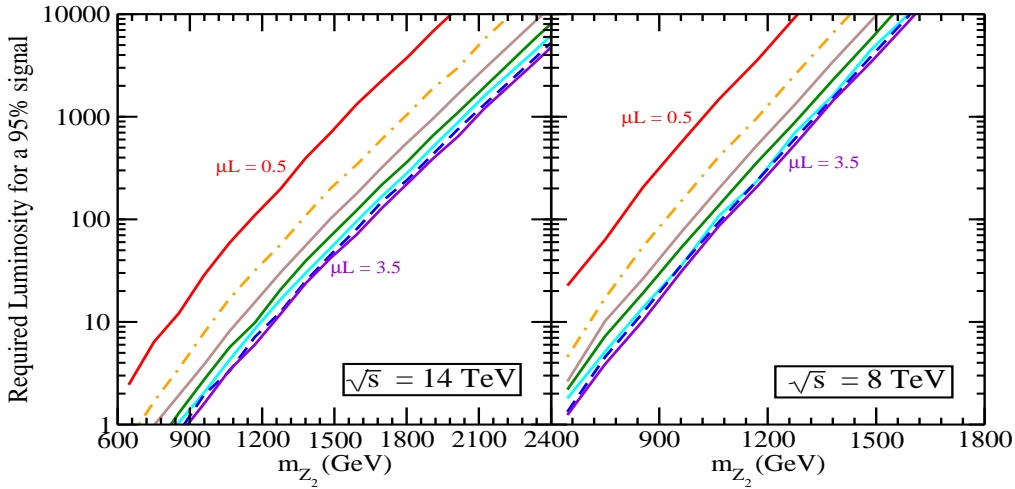}
\caption{\em Luminosity required to have $95\%$ signal over
  background at the LHC.}
  \label{f:lumi95}
\end{figure}

In Fig.-\ref{f:lumi95}, we have shown the variation of the required
luminosity at LHC to have signal event with at least $95\%$ confidence
level, defined as $\sigma = \sqrt{S/(S+B)}$, with the $Z_2$ mass for a
different set of values of the bulk mass parameters.  We can see from
the figure that our analysis can probe upto a few TeV of $Z_2$ mass
with  $\sim 100~ {\rm fb}^{-1}$ luminosity at the LHC.

\section{Conclusions}

The Standard Model is almost complete with the discovery of the new
boson at the LHC experiment. Beside some theoretical issues the signal
is also not in complete agreement with what we expect from SM. These
issues keep LHC to continue to its new physics search beyond the
standard model. LHC has now reached to an energy scale and will
continue to gear up more either to discover or turned down many new
physics scenario. At this energy  scale it produces quite high
energetic multiple number jets with hard leptons. LHC has discussed  
the $Z$ plus jets as a final state to test different SM issues. It is
thus timely to have a look at some of the new physics in this
directions. 

At the tree level a single jet plus dilepton may appear in a few new
physics scenario, split-UED is one of such scenario. There is always a $Z_2$-like
symmetry associated with most of the new physics, where a heavy stable
particle is escape collider as missing energy. However, in split-UED the
heavy weak neutral gauge bosons can directly decay into a pair leptons.

We consider in this paper the production of such a heavy weak gauge
boson associated with a hard jet. Here our analysis is at the parton level
and so a further detail analysis is much needed. As the heavy gauge
boson will directly decay into a pair of leptons, they will have a
very large momentum. This high energetic jet and leptons
allow us to use a strong cut on the $p_T^{\rm jet}$ and the dilepton
invariant mass greater than $0.9$ times the $\gamma_2$ mass. In
addition, we have also used a rapidity cut $|\eta| < 2.5$ for leptons
and consider the leptons which are isolated from jet by $\Delta R >
0.7$ only. We can have various SM channel to mimic our
signal. However, the strong cut applied here can wash out any such
background. Our analysis can probe upto a few TeV of heavy boson mass
with a $\sim 100 {\rm fb}^{-1}$ luminosity at LHC.

\acknowledgments
This work is supported by Basic Science Research Program through the
National Research Foundation of Korea funded by the Ministry of
Education, Science and Technology (2011-0010294) and (2011-0029758).

\end{document}